\documentclass[10pt,letterpaper]{article}
\usepackage{color}
\usepackage{graphicx}

\usepackage{ae} 

\title{Offset frequency dynamics and phase noise properties of a self-referenced 10 GHz Ti:sapphire frequency comb}
\author{Dirk C. Heinecke\textsuperscript{1,2,*}, Albrecht Bartels\textsuperscript{3}, and Scott A. Diddams\textsuperscript{1,+}}

\begin{document}

\maketitle
\textsuperscript{1}National Institute of Standards and Technology, 325 Broadway M.S. 847, Boulder,
CO 80305, USA \\
\textsuperscript{2} Center for Applied Photonics, University of Konstanz, Universit\"{a}tsstrasse 10,
78457 Konstanz, Germany\\
\textsuperscript{3} Gigaoptics GmbH, Blarerstrasse 56, 78462 Konstanz, Germany\\

* dirk.heinecke@uni-konstanz.de, + sdiddams@nist.gov
\begin{abstract}
This paper shows the experimental details of the stabilization scheme that allows full control of the repetition rate and the carrier-envelope offset frequency of a 10 GHz frequency comb based on a femtosecond Ti:sapphire laser. Octave-spanning spectra are produced in nonlinear microstructured optical fiber, in spite of the reduced peak power associated with the 10 GHz repetition rate. Improved stability of the broadened spectrum is obtained by temperature-stabilization of the nonlinear optical fiber. The carrier-envelope offset frequency and the repetition rate are simultaneously frequency stabilized, and their short- and long-term stabilities are characterized. We also measure the transfer of amplitude noise of the pump source to phase noise on the offset frequency and verify an increased sensitivity of the offset frequency to pump power modulation compared to systems with lower repetition rate. Finally, we discuss merits of this 10 GHz system for the generation of low-phase-noise microwaves.
\end{abstract}

\section{Introduction}
Frequency-comb sources with mode spacing in the range of tens of gigahertz are of increasing interest for applications in optical frequency metrology, microwave photonics \cite{Huang2008}, photonic microwave generation \cite{Bartels2005}, and astronomical \cite{Li2008} and laboratory spectroscopy \cite{Diddams2007}. In all such applications, the stabilization of the mode frequencies is required. The frequency of the n$th$ frequency comb mode is given by the comb equation $\nu_n= n \cdot f_R + f_0$, where $f_R$ is the frequency spacing between comb lines given by the laser repetition rate and $f_0$ is the carrier-envelope offset frequency, see e.g. \cite{Cundiff2002}. Therefore stabilization of the mode frequencies requires control over both degrees of freedom, repetition rate and offset frequency. There are many promising approaches to frequency-comb sources with multiple gigahertz mode spacing, including  active and passive mode-locking of solid-state or fiber lasers \cite{Quinlan2006,Yoshida1998,Paschotta2004,Zeller2007,Carruthers1996,Martinez2011}, cavity filtering \cite{Quinlan2010,Kirchner2009,Steinmetz2009}, microcavities \cite{Del'Haye2008,Savchenkov2008,Braje2009} and electro-optic modulation \cite{Xiao2009}. However, for many of the laser-based approaches, the pulse energy is too low to provide an octave-spanning spectrum, as required for self-referenced stabilization of the offset frequency \cite{Telle1999,Jones2000}. With narrower-bandwidth comb sources, it is possible to use optical spectroscopy and higher-order nonlinear optics \cite{Heinecke2009}, or auxiliary sources \cite{Del'Haye2008} to stabilize $f_0$, although this adds complexity and removes the benefit of stabilization to well-established microwave references such as commercially available atomic clocks or the global positioning system (GPS). \\
In a recent report \cite{Bartels2009} we demonstrated a self-referenced octave-spanning spectrum produced from a 10 GHz mode-locked Ti:sapphire laser.  This was accomplished with only 50 pJ pulse energy available for nonlinear broadening, but taking advantage of the fact that the laser produced $\sim$ 30 fs pulses to provide a peak power of 1.25 kW. Here we provide complete details of the stabilization technique, characteristics of the stabilized signals, and our efforts aimed at improved long-term stability.\\
In our approach, we use an end-sealed and temperature stabilized microstructured fiber. This allows stable generation of an octave-spanning spectrum and reliable locking of the comb with stable operation for times scales of 1 to 2 hours.  The residual phase noise on $f_0$ is measured and the frequency stability of the comb is characterized in both the free-running and the locked case. The achieved frequency stability for time scales greater than 1 s is limited by the employed microwave references. In addition, we examine the transfer of amplitude noise on the pump laser to phase noise on the offset frequency. For systems of high repetition rate this phase noise is expected to be higher than that for a comparable system with lower repetition rate.  Finally, the unique properties of the laser, including its high repetition rate, short pulses and high average power, make it an interesting candidate for high-power low-noise microwave generation. Along these lines, we demonstrate the generation of a 10 GHz microwave signal with 0 dBm power directly from a high-speed, high-power InGaAs photodiode that detects the 10 GHz pulse train.

\section{Generation of an octave-spanning continuum}
\begin{figure}
\centering
\includegraphics[width=0.75\textwidth]{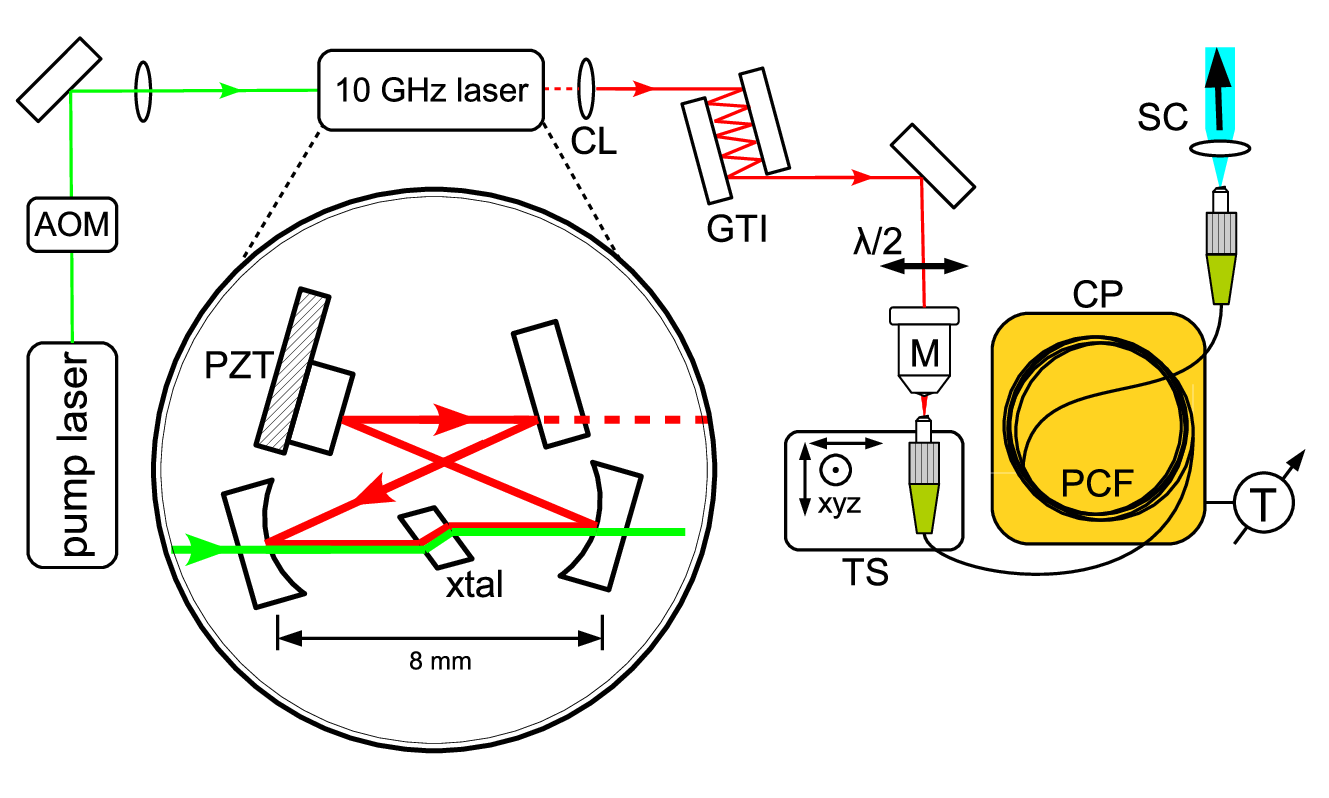}
\caption{Compact setup for the supercontinuum generation. The zoom in shows the 10 GHz Ti:sapphire laser cavity. AOM: acousto-optic modulator, PZT: piezo-electric transducer to modulate cavity mirror, CL: collimation lens, GTI: negative dispersion Gires-Tournois interferometric mirrors for pulse recompression, $\lambda$/2: half-wave plate, M: 60x microscope objective, TS: xyz-translation stage, PCF: 1.5 m of microstructured fiber, CP: temperature-stabilized copper plate, SC: generated supercontinuum.}
\label{fig:laser_pcf}
\end{figure}

The Ti:sapphire laser used in this work is passively modelocked via the Kerr lens effect and consists of a four-mirror ring cavity with a 30 mm round-trip length corresponding to a repetition rate of 10 GHz. While pumping with 8.5 W, the laser output power is about 1 W and the pulse length is below 30 fs. Due to the high repetition rate the energy per pulse is only 100 pJ. More details about the laser can be found in Ref. \cite{Bartels2008}. The inset of Fig. \ref{fig:laser_pcf} shows the schematic of the laser together with the setup for the supercontinuum generation.

\begin{figure}
\centering
\includegraphics[width=0.85\textwidth]{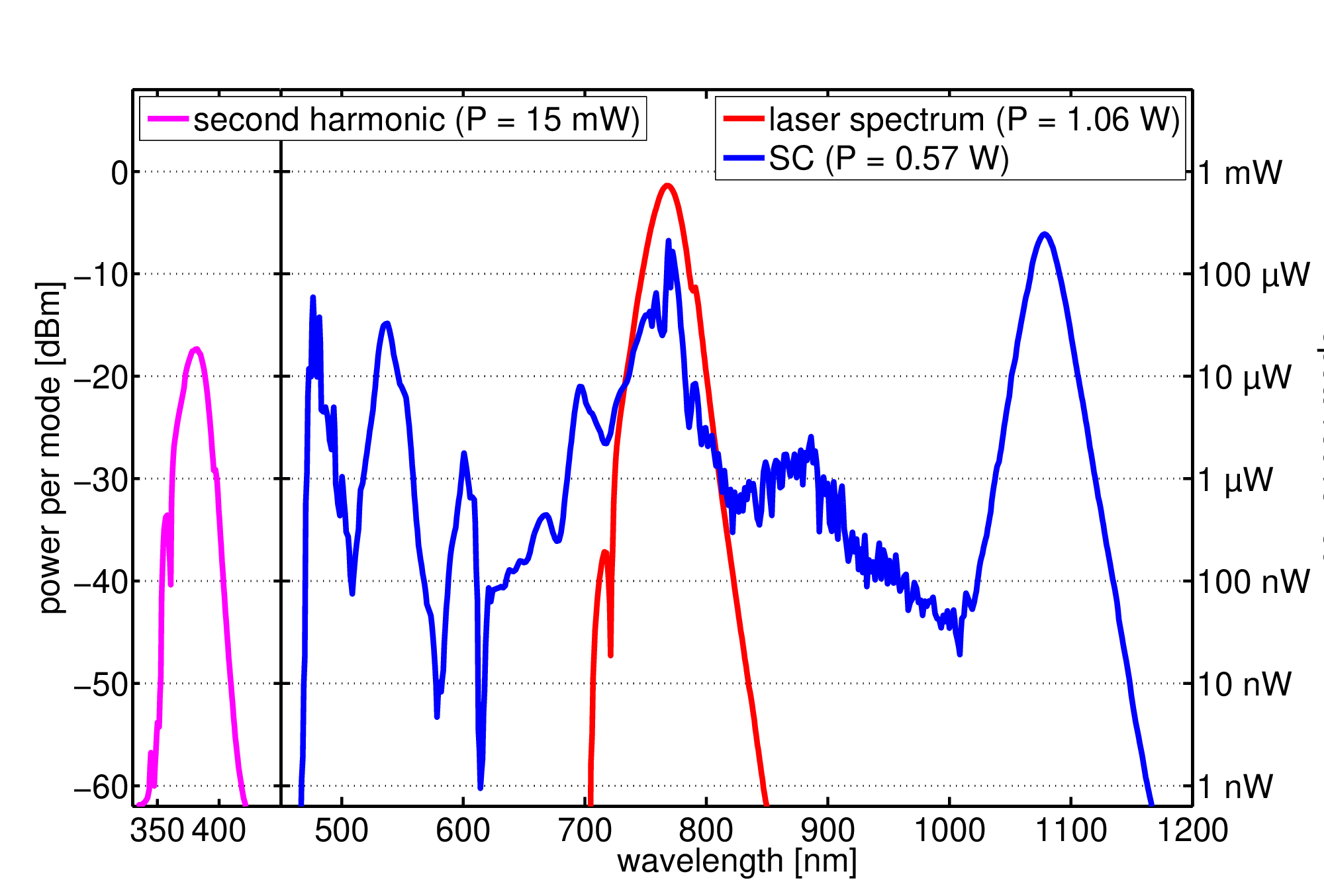}
\caption{Laser spectrum and octave-spanning spectrum (supercontinuum) generated in microstructured fiber. To further demonstrate the possibility of expanding the spectral coverage of the comb, the spectrum of the laser doubled in a 50 $\mu$m thick BBO crystal is also shown.}
\label{fig:laserspectrum}
\end{figure}

We use an acousto-optic modulator in the pump beam for pump power control. For repetition rate tuning, one of the cavity mirrors is mounted on a piezoelectric transducer. In order to achieve sufficient nonlinear broadening in the microstructured fiber, the peak power of the laser pulses coupled into the fiber has to be as large as possible. Therefore the pulses need to be as short as possible and the coupling efficiency has to be maximized.
To compensate for the fiber coupling optics the laser output is recompressed and prechirped by a pair of negative-dispersion Gires-Tournois interferometric mirrors. The collimated light is focused with a microscope objective into a commercially available microstructured fiber. The fiber core has a diameter of 1.5 $\mu$m and the zero-dispersion wavelength of the fiber is at 590 nm. The microstructure is end-sealed and the fiber is connectorized and angle polished. This allows a more robust coupling compared to that of a bare fiber tip. The coupling efficiency can exceed 50 \%. For a coupled power of 550 mW the peak power is about 1.8 kW. With a half-wave plate the polarisation is adjusted for maximum broadening in the microstructured fiber. The supercontinuum spectrum is shown in Fig. \ref{fig:laserspectrum}. 
In early experiments \cite{Bartels2009} the fiber-coupling stability limited the operation of the system to intervals of about 10 minutes. The large average power in combination with tight focussing into the small fiber core leads to heating of the fiber and thermal instabilities in coupling efficiency and output polarization. To improve the stability of the supercontinuum, the fiber is thermally connected to a temperature-stabilized copper plate. A slow feedback loop keeps the temperature of the plate and the fiber constant within 5 mK. After coupling light into the fiber and a thermalization period of about one hour, stable coupling over several hours can be achieved.

\section{Measurement and stabilization of the offset frequency}\label{sec:fodetect}
\begin{figure}
\centering
\includegraphics[width=0.8\textwidth]{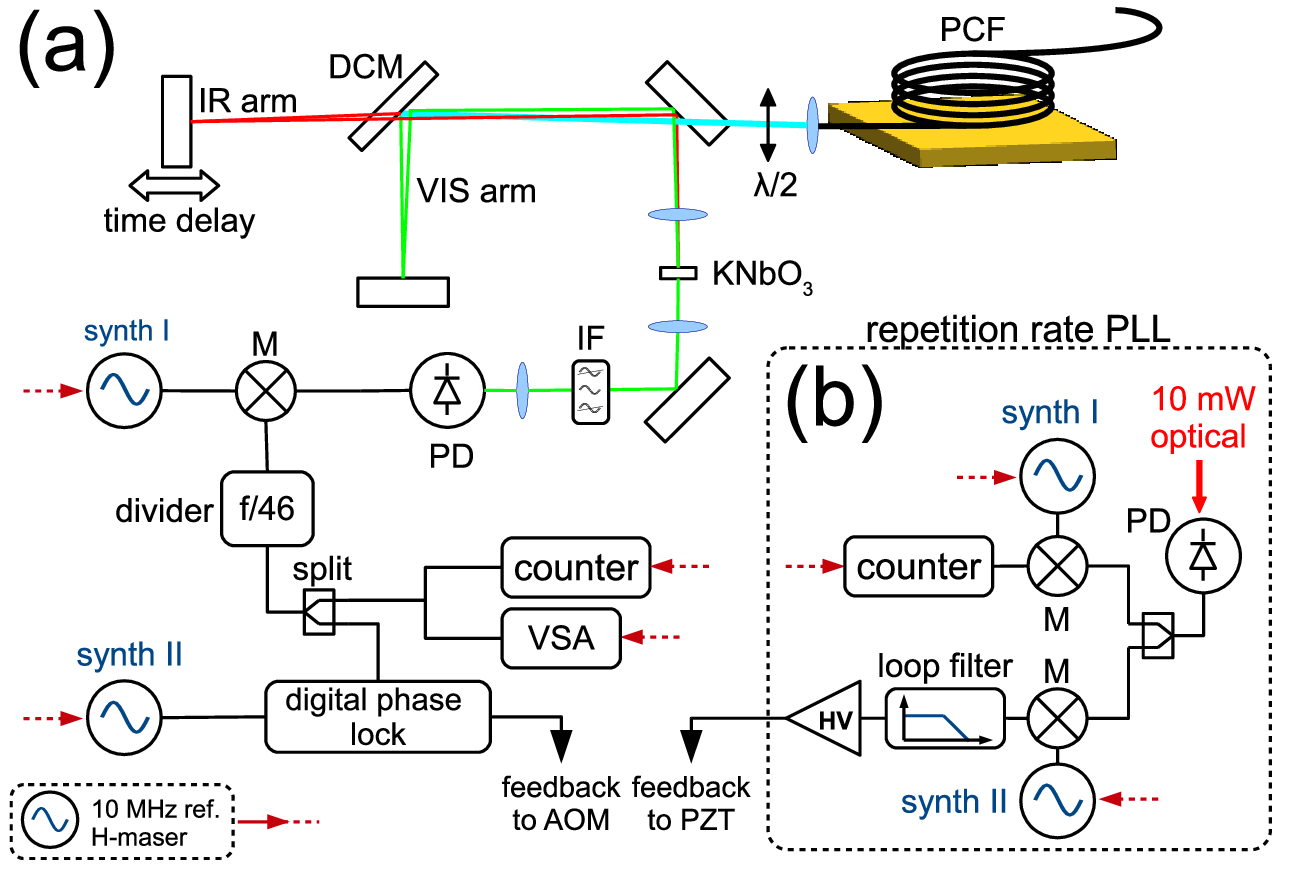}
\caption{(a) Setup for offset frequency detection, locking and analysis consisting of the f-2f interferometer and electronics for signal processing. PCF: microstructured fiber on a temperature-stabilized copper plate, DCM: dichroic mirror, KNbO$_{\mathrm{3}}$: 1\,mm long nonlinear potassium niobate crystal, IF: interference filter as optical bandpass at 532\,nm with 3\,nm bandwidth, PD: SI PIN diode with 2\,GHz bandwidth, M: double-balanced mixer, synth I: synthesizer at $f=f_0$ + 925\,MHz, div: frequency divider with N~=~46, split: -3\,dB/-3\,dB signal splitter, VSA: vector signal analyzer. For clarity signal microwave filters and amplifiers are not shown. (b) Repetition rate phase-locked loop and characterization scheme. PD: high-speed photodetector with 10\,GHz bandwidth, M: double-balanced mixers for phase detection and frequency down-conversion, synth I: synthesizer at $f=f_R$+10 MHz, synth II: synthesizer at $f=f_R$.}
\label{fig:f2finterferometer}
\end{figure}

For offset frequency detection we use the f-2f scheme requiring an octave-spanning spectrum. The Michelson interferometer setup, including the signal-processing electronics for offset frequency stabilization, is shown in Fig. \ref{fig:f2finterferometer}. Light around 1064 nm is frequency-doubled in a nonlinear potassium niobate crystal (KNbO$_3$). To obtain interference with directly generated 532 nm light, both beams have to overlap in wavelength, space, time, and polarization. Since the different wavelengths have different divergence, only parts of the spectrum are well collimated. The dichroic mirror separates the 532 nm part of the continuum in one arm of the interferometer, while light at 1064 nm goes into the second arm. The delay in the second arm is used to adjust the time delay between the interfering spectral components. With a wave plate at 1064 nm the polarization of the IR light is rotated to maximize the frequency doubling efficiency. With a 532 nm bandpass filter, only the interfering parts of the spectrum are selected before being focused on the photodetector. The photodetector is a Si PIN diode with a cutoff frequency above 2\,GHz. For most measurements the offset frequency was in the range of 2.7\,GHz. After amplifying, the signal is bandpass filtered and mixed down to 925\,MHz by using a double-balanced mixer, where the LO port is driven by a synthesizer at a frequency of about 3.7\,GHz. After a second amplification step the frequency is divided by a factor of N = 46 down to 20\,MHz \cite{NOTE}. This signal is used as an input for the offset-frequency phase detector. For characterization we measure the offset frequency with a frequency counter, and the phase-noise power spectral density with a vector signal analyzer. All synthesizers and counters are referenced to the same 10\,MHz hydrogen maser signal via a distribution amplifier. To stabilize the offset frequency we use a phase-locked loop based on a digital phase detector \cite{AnalogDevices}. 

\begin{figure}
\centering
\includegraphics[width=\textwidth]{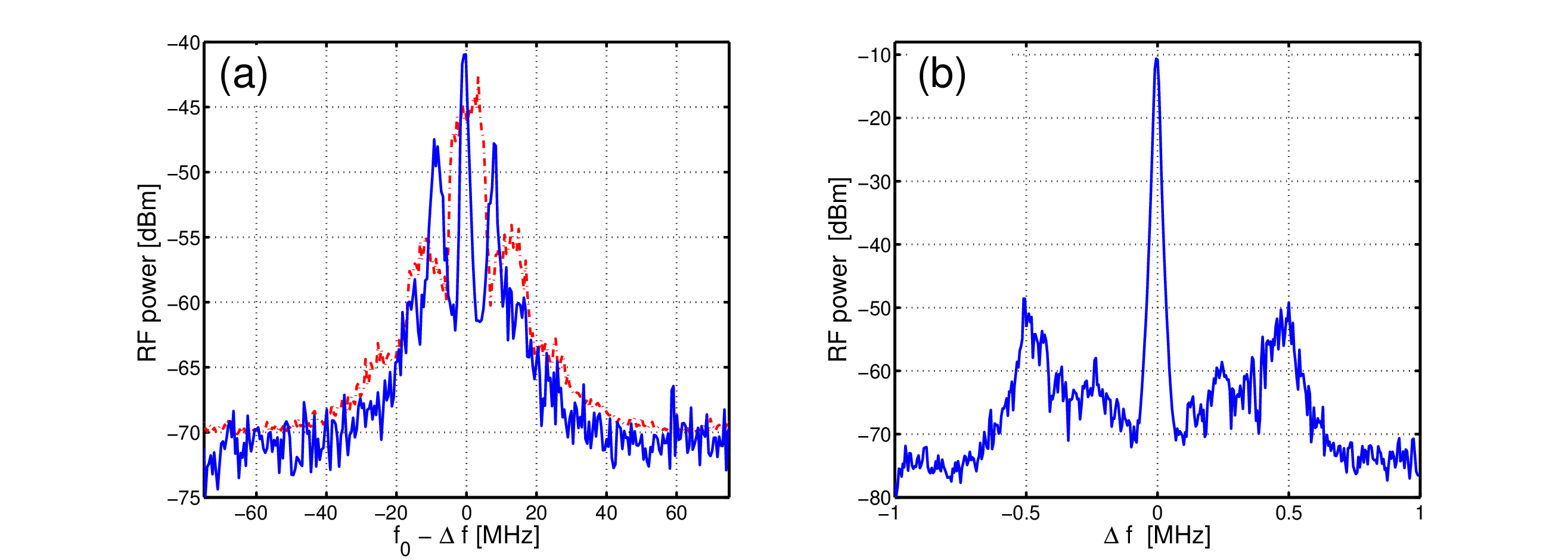}
\caption{Offset frequency beat signal for the free-running system. (a) Microwave spectra of the unlocked and undivided offset frequency signal at 2.7 GHz taken with 1 MHz resolution bandwidth. The red trace (\textcolor{red}{- -}) shows a spectrum where the pump laser exhibits excessive amplitude noise. (b) High-resolution zoom (10\,kHz resolution bandwidth) in the spectrum of the mixed and divided signal of the offset frequency at the carrier frequency of 20 MHz.}
	\label{fig:f0beat}
\end{figure}

Figure \ref{fig:f0beat} (a) shows spectra of the offset frequency measured with a microwave spectrum analyzer for the free-running system. Sidebands at about 8\,MHz are clearly visible. In addition, Fig.\,\ref{fig:f0beat} (a) shows a spectrum where the femtosecond laser shows amplitude noise. This is an intermittent phenomenon linked to amplitude noise on the pump laser. The sidebands visible in Fig. \ref{fig:f0beat} (a) have approximately the same spacing of 8 MHz as sidebands generated in Q-switched operation of the laser. Interestingly, these sidebands are not evident in the direct microwave spectrum of the repetition rate. However, when the laser is made to Q-switch by misalignment, sidebands are clearly seen in the microwave spectrum, and the many high-order harmonics of 8\,MHz appear in the offset frequency spectrum. Clearly the offset frequency is very sensitive to amplitude fluctuations. Figure \ref{fig:f0beat} (b) shows a 2\,MHz zoom into the offset frequency spectrum. The noise on the offset frequency carrier is further addressed in Sec.\,6.

\section{Stabilization of the repetition rate frequency}
To stabilize the repetition rate, we use about 10\,mW of light that is reflected from the wave plate in front of the microstructured fiber and focus it on a high-speed photodetector. The schematic setup is shown in Fig. \ref{fig:f2finterferometer} (b). The 10\,GHz microwave signal is amplified and split. One part goes to the phase-locked loop, while the other part is used for analysis. The latter signal is mixed down to 10 MHz and can then be counted. The PLL employs a double-balanced mixer for phase detection between repetition rate and reference synthesizer. A loop filter provides low-pass filtering and adjustable gain. The error signal is fed to a high-voltage amplifier driving the piezoelectric transducer carrying an intracavity mirror.

\section{Frequency stability analysis}
\begin{figure}
\centering
\includegraphics[width=\textwidth]{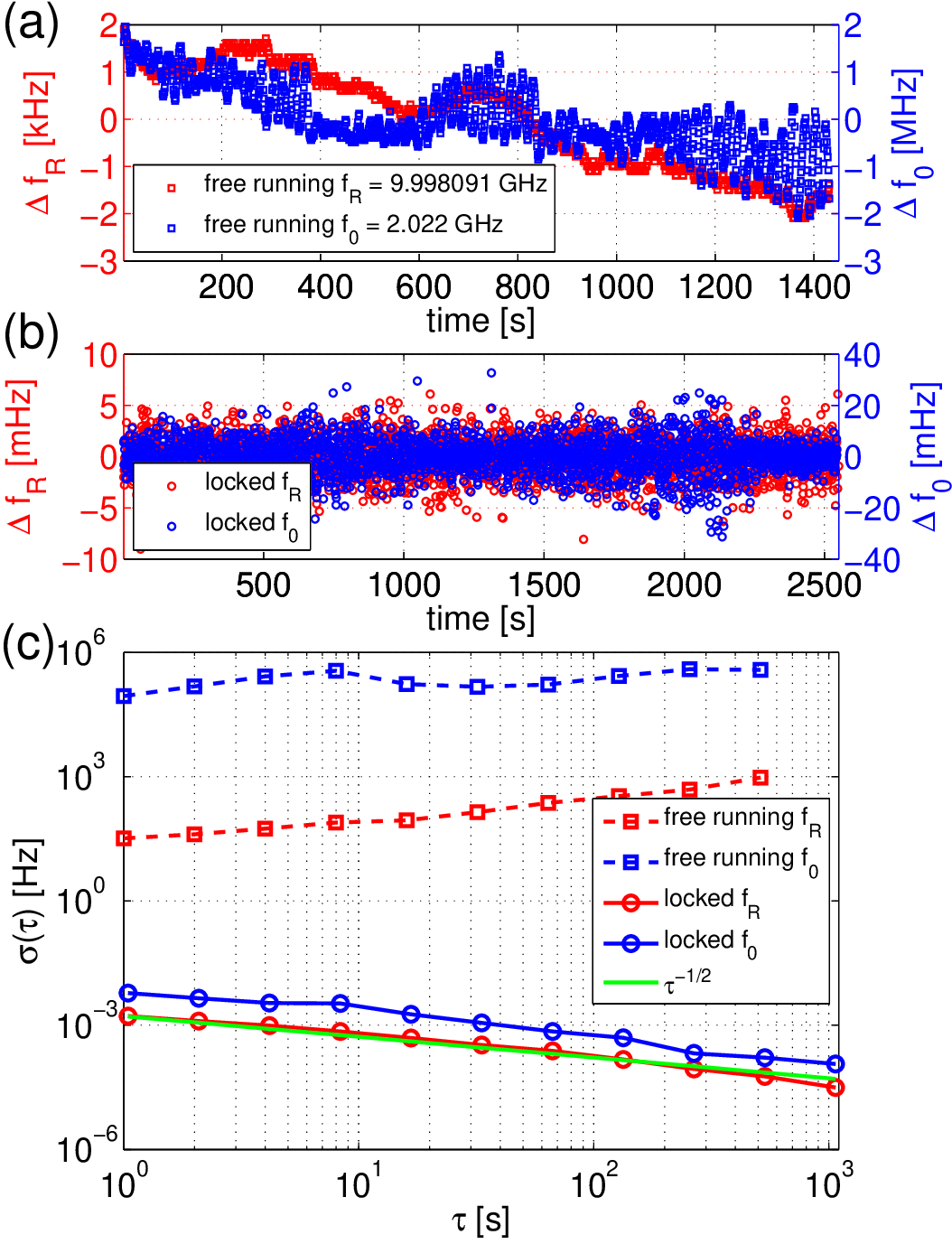}
\caption{(a) Counting record of the offset frequency and the repetition rate for the unlocked system with 1 s gate period. The vertical coordinate axis shows the changes in frequencies. (b) Counting record of the offset frequency and the repetition rate for the locked system with 1 s gate period. c) Calculated Allan deviation for both frequencies. As a guide to the eye, the green line marks a $\tau^{-\frac{1}{2}}$ dependency. The offset frequency data are corrected for the division factor and represent the stability of the actual offset frequency.}
\label{fig:f0fR_CR_Allan}
\end{figure}

Both, the divided offset frequency and the mixed-down repetition rate can be measured using frequency counters. In all counter measurements we use a 1 s gate period. Figure~\ref{fig:f0fR_CR_Allan}~(a) shows the behaviour of the free-running system (all phase locks off) over a time span of half an hour.  Both frequencies show a large long term drift, where the change of the offset frequency is more than three orders of magnitude larger than the drift of the repetition-rate frequency. The frequency drift can be attributed to temperature changes of the room and the laser base plate. Figure \ref{fig:f0fR_CR_Allan} (b) shows the offset frequency and repetition rate of the locked system counted over a period of 45 minutes. The system stays locked for about one and a half hour, before adjustments to the phase locks become necessary to continue operation. The calculated Allan deviations for the free-running and the locked case are shown in Fig. \ref{fig:f0fR_CR_Allan} (c). As described in Fig.\,\ref{fig:f2finterferometer} (a) and (b) four synthesizers are employed: two as references for the phase-locked loops and two for mixing down the frequencies. The stability of the beat signal is then a combination of the stability of both synthesizers and is dominated by the synthesizer that is less stable. The measured fractional instability at one second of the repetition rate and the offset frequency are
\begin{eqnarray}
\sigma_{f_R}(1\,s) = \frac{\delta f_R}{f_R} = \frac{1.9 \cdot 10^{-3} \mathrm{Hz}}{10 \cdot 10^9 \mathrm{Hz}} = 1.9\cdot 10^{-13} \label{gl:ADEVfR}\\
\sigma_{f_0}(1\,s) = \frac{\delta f_0}{f_0} = \frac{6 \cdot 10^{-3} \mathrm{Hz}}{2.7 \cdot 10^9 \mathrm{Hz}} = 2.2\cdot 10^{-12}.
\end{eqnarray}
This corresponds to the fractional instability provided by the synthesizers (which we have
characterized separately) and the hydrogen maser. Thus, the stability is presently limited by the synthesizers and not by the laser or the phase-locked loops. The Allan deviation for both degrees of freedom follows the $\tau^{-\frac12}$ behavior expected for white phase noise, considering a small dead period in the counting process. These data indicate that the repetition rate is ultimately limited by the maser reference, while for the offset frequency the synthesizers add noise and therefore limit the stability to about $2.2\cdot 10^{-12}$ at 1\,s. We can expect to improve the offset frequency stability by using better synthesizers.

An estimate of the optical stability can be made by use of the values for the microwave frequencies $f_R$ and $f_0$ and projecting them into the optical domain via the comb equation. We get a fractional instability of $ 1.9  \cdot 10^{-13}$ at 1\,s for the repetition-rate contribution and $1.55 \cdot 10^{-17}$ at 1\,s for the offset-frequency contribution. 

\section{Laser amplitude noise to offset frequency phase-noise conversion}
While counting a frequency over long periods measures how well the line center is determined depending on the averaging period, a phase-noise analysis gives information of fluctuations on time scales less than 1\,s. A low-phase noise is particularly important for the generation of low-phase-noise microwave signals and for metrology applications. For the phase-noise analysis, a vector signal analyzer is used. The vector signal analyzer demodulates the signal at the carrier frequency and then measures the power spectral density of the phase noise. The spectral density of phase fluctuations is measured for the divided offset frequency signal at 20\,MHz. Since phase noise scales in a PLL with the carrier frequency, the actual noise density on the offset frequency carrier is recovered by multiplying the measured spectral noise density by the square of the division factor. Integrating from 10\,Hz to the cut-off frequency of the spectrum analyzer of 3\,MHz gives the integrated or accumulated phase noise over that bandwidth.

\begin{figure}
	\centering
		\includegraphics[width=1\textwidth]{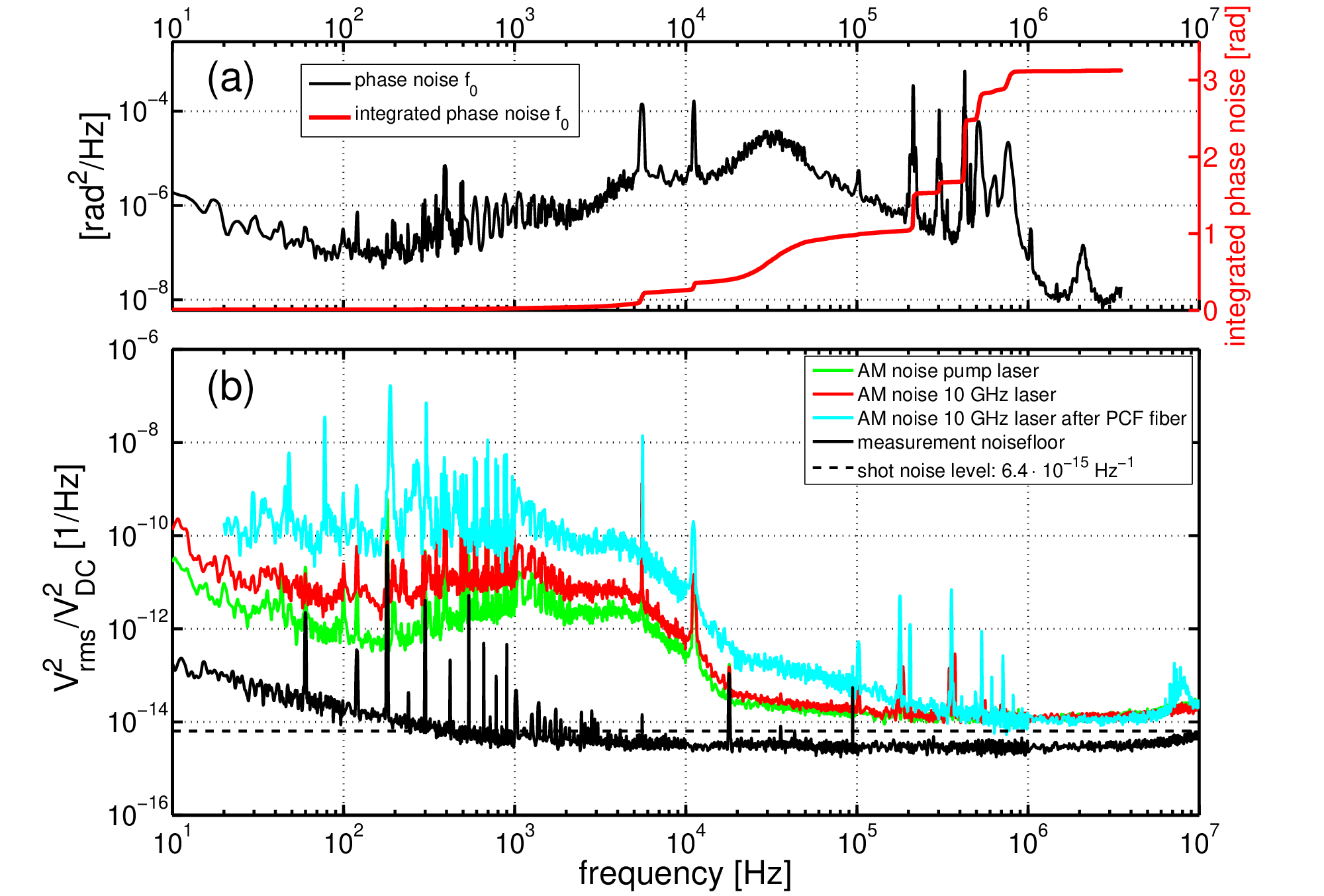}
		\caption{(a) Offset frequency phase noise and integrated phase noise. (b) Amplitude noise of the pump laser and the 10 GHz laser and corresponding noise floors. The measured integrated relative intensity noise for the pump laser and the 10 GHz laser is 0.047 \%  and 0.054 \%, respectively.}
	\label{fig:noisePMAM}
\end{figure}

Figure \ref{fig:noisePMAM} (a) shows the phase noise measurement. The bandwidth of the f$_0$ PLL is about 18 kHz, as evident by the peak in the spectrum near this frequency. The integrated phase noise for this measurement is about 3.1 radians. To reduce the phase noise further requires improving the PLL. At low frequencies up to the kilohertz range the noise contributions have thermal, mechanic and electronic origins, but their frequencies are within the servo bandwidth. For higher frequencies the loop gain is reduced due to the limited bandwidth of the AOM and phase shifts or delays in the loop. 

It is interesting to note that the integrated phase noise of three radians is consistent with measurements on the f$_0$ phase noise of a similar ring Ti:sapphire laser operating at 1 GHz repetition rate. In that case, we have obtained an integrated noise that is about 0.32 rad over a comparable bandwidth. The 10-times increase in going to the 10 GHz system is to be expected if laser parameters (other than the reduced air path) are approximately the same. This is clear if we consider that the carrier-envelope phase shifts are similar in magnitude, but is multiplied by the repetition rate according to

\begin{equation}
f_0(t) = \frac{\Delta \Phi_{CE} (t)}{2 \pi} \cdot f_{R},
\end{equation}
where $\Delta \Phi_{CE}$ specifies the carrier-envelope phase shift (modulo 2$\pi$) of a cavity round-trip (see e.g. \cite{Cundiff2002}). Thus, the repetition rate of the system should also be considered when comparing the integrated carrier envelope phase noise of different frequency-comb systems.

In order to determine noise sources and contributions to the f$_0$ phase noise, we performed a number of amplitude noise measurements at different positions in the system. A major contribution to the phase noise is expected to be the pump laser. To measure amplitude noise (AM) a portion of the light from the pump laser or the 10 GHz laser is focused onto a 125 MHz photodetector and the signal is sent to a vector signal analyzer. The signal analyzer measures a noise power spectral density from 0.6\,Hz up to 10\,MHz. Figure \ref{fig:noisePMAM} (b) summarizes these measurements. 
All measurements including the noise floor measurement show a strong 60\,Hz peak and higher harmonics of this peak originating from pick-up from the power line. It is most likely that AM noise on the pump laser with frequencies higher than 100\,kHz causes the broadening of the offset frequency shown in Fig.\,\ref{fig:f0beat}. Comparison with the calculated shot noise level shows that there is a shot noise contribution to the noise density for frequencies higher than 10 kHz.
Comparing the amplitude noise of the pump and the 10 GHz laser shows a direct transfer of the AM noise. This is consistent with a model discussed by Scott et al. \cite{Scott2007} for the AM noise transfer, which predicts uniform AM-to-AM noise transfer up to the relaxation oscillations frequency, which is about 3 MHz in our case. The transfer occurs via changes in gain dynamics, where a change of the pumping rate changes the cavity photon number. Above the relaxation oscillations the level population can no longer follow the changes of the pump rate, leading to a suppression of the noise transfer. Most amplitude noise peaks show up as phase noise on the offset frequency. Again, this is because AM noise also transfers to phase noise up to the relaxation frequency. We also measure the amplitude noise after the microstructured fiber, and can see that the main effect is an amplification of the noise spectrum in the microstructured fiber. The frequency dependent amplitude-to-phase noise conversion in microstructured fiber has been shown \cite{Fortier2002} to scale linearly with modulation frequency. In addition, there is a peak at 8\,MHz, the same frequency where we see large sidebands on the offset frequency beat. This amplitude noise analysis confirms the transfer from pump laser amplitude noise to offset frequency noise.

\section{Prospects on shot-noise limited microwave generation}
Finally we consider the properties of the 10 GHz laser presented here to the generation of ultralow phase noise microwave signals. We have introduced an approach that uses the frequency comb to transfer the exceptional instability of optical frequency references(e.g. $\sim 10^{-15}$ \cite{Young1999}) to the microwave domain.  In this case, the stable optical frequency
reference is employed to stabilize the comb modes in the optical domain, and a microwave signal at the comb repetition rate is generated by detecting the pulse train with a photodetector. This technique has been demonstrated for frequency combs with repetition rates up to 1 GHz, and the desired microwave frequency is obtained via harmonic generation in the photodetector (e.g., 10 GHz signal obtained from the 10${th}$ harmonic of a 1 GHz laser) \cite{Bartels2005,Fortier2011}. The close-to-carrier phase-noise properties of the generated microwave signal depend on several factors, including the phase noise on the reference and free-running comb, as well as the noise level of the various servo loops involved.  However, for solid-state laser sources, the limiting high-frequency noise floor on the microwave signal is typically provided by the shot noise on the photodetected repetition-rate signal. In the ideal case, the shot-noise power increases linearly with the incident optical power, while the microwave signal power scales quadratically with optical power. Thus, the signal-to-noise ratio of the 10 GHz microwave signal increases linearly with the detected light power.  However, the onset of saturation in the photodetection process limits the amount of signal power that can be generated.  This is a particular problem when we consider the detection of high harmonics of the repetition rate. Previously, we have reduced the impact of photodiode saturation by optical filtering a lower-repetition-rate source with a Fabry-Perot cavity \cite{Diddams2009}; however, the use of a self-referenced and optically-stabilized 10 GHz frequency comb would eliminate the complication of filtering when 10 GHz signals, or even higher frequencies, are desired. 

\begin{figure}
	\centering
		\includegraphics[width=0.75\textwidth]{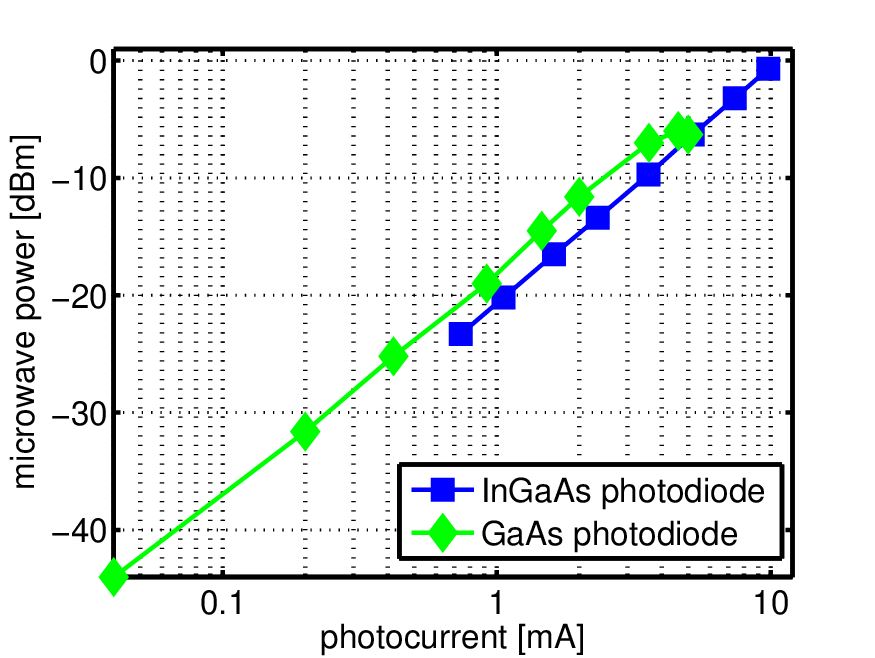}
	\caption{Microwave power at 10 GHz versus photo current for two different photodetectors. Both photodetectors were terminated with 50 $\mathrm{\Omega}$ resistors.}
	\label{fig:microwaveplot}
\end{figure}

To evaluate the potential shot-noise floor of signals generated from the 10 GHz laser, we have detected the repetition rate with two different high-speed photodiodes. Figure \ref{fig:microwaveplot} shows the achieved microwave power at 10 GHz versus photo current for GaAs and InGaAs PIN diodes. While the GaAs photodiode has improved responsivity, the 10 GHz power saturates above 3 mA of average photocurrent. However, in the case of the InGaAs photodiode, we see no evidence of saturation, even at photocurrents of 10 mA, which provides 0 dBm of microwave power.  This power level is similar to what we were able to achieve with a 5 GHz pulse train produced by a filtered 1 GHz source \cite{Diddams2009}, and should support a shot-noise-limited floor of -162 dBc/Hz.

\section{Conclusion}
In this paper we present a detailed discussion of the stabilization of both degrees of freedom of a 10 GHz Ti:sapphire frequency comb. A challenge for implementing a self-referencing scheme for a high-repetition-rate system is the intrinsic low energy per pulse available to drive the nonlinearity of the microstructured fiber. We overcome this issue by using a small-core microstructured fiber that is temperature stabilized to prevent thermal instabilities. This allows the generation of a spectrum spanning a full octave. The offset frequency and the repetition rate can be locked simultaneously by use of standard phase-locked loop techniques. The long-term stability is characterized and we find a stability of the comb that is limited by the hydrogen maser reference. We also measure the transfer of amplitude noise of the pump source to phase noise on the offset frequency and find an increased sensitivity of the offset frequency to pump power modulation compared to that of lower-repetition-rate systems. We also demonstrate the potential of this system for low-noise microwaves at 10 GHz. The high repetition rate allows microwave powers up to 0 dBm without saturating the photodiode. \\
With further technical improvements the 10 GHz frequency comb can be used as a reliable tool for metrology with a performance that is comparable to that of existing systems. In frequency-comb applications such as astronomical spectrograph calibration as well as mode-resolved spectroscopy, multichannel communication and line-by-line pulseshaping a high repetition rate is beneficial. Here, the 10 GHz frequency comb could prove to be an excellent comb source with a stabilization limited only by its reference, and providing a high repetition rate in combination with femtosecond pulses and high average power.

\section{Acknowledgements}
We thank F. Quinlan and T. Fortier for experimental assistance and helpful discussions, A. Joshi and S. Datta of Discovery Semiconductor for providing the 10 GHz InGaAs photodiode and  M. Kirchner and G. Ycas for their comments on this manuscript. This work was supported by NIST, an agency of the US government, and is not subject to copyright in the US.

\bibliographystyle{osajnl}

\begin{thebibliography}{99}
\bibitem{NOTE}
For typical frequency counters the input frequency is limited to a few hundreds
  of megahertz. Thus, the frequencies to analyze in the range of gigahertz have
  to be converted to lower values to enable counter measurements. In the
  process of difference or sum frequency generation in a mixing device the
  actual frequency fluctuations of the input frequencies are preserved. Using a
  stable reference frequency this allows to shift frequencies without affecting
  the frequency stability. In contrast, in a frequency division process the
  noise density and therefore the frequency fluctuations are reduced by the
  division factor. This leads to an increase in frequency stability. For the
  offset frequency stabilization and analysis we use a frequency division step
  to reduce phase fluctuations making the phase lock more robust by extendig
  the capture range of the phase detector. In the data evaluation we account
  for this division factor to recover the original offset frequency stability.
\end{thebibliography}



\end{document}